\documentclass[11pt]{article}
\usepackage{hyperref}
\pdfoutput=1

\begin{document}

\title{Liquid-solid impacts of yield-stress fluids}

\author{Marc E. Deetjen, Brendan C. Blackwell, \\Joseph E. Gaudio, Randy H. Ewoldt \\
\\\vspace{6pt} Department of Mechanical Science \& Engineering, \\ University of Illinois at Urbana-Champaign\\ Urbana, IL, 61801, USA}

\maketitle


\begin{abstract}
This is an entry to the Gallery of Fluid Motion at the 66th annual meeting of the APS-DFD, held November 2013 in Pittsburgh, PA. In this fluid dynamics video we demonstrate distinct features of yield-stress fluid droplets impacting pre-coated surfaces.
\end{abstract}


\section{Description of video}

Yield-stress fluids exhibit a distinct type of non-Newtonian behavior: they are effectively fluid at high stress and solid at low stress. When droplets of yield-stress fluids impact pre-coated surfaces, they exhibit splashing behavior that is not observed in impacts onto dry surfaces. We use droplets of an aqueous solution of Carbopol 940 (a shear-reversible physical gel with 0.1 - 1.0wt\% additive). We explore how the characteristics of these impacts vary as a function of three input parameters. First, we vary the material properties by changing the concentration of the additive, including Newtonian droplets for comparison (water and glycerol). We then vary the droplet size and the impact velocity. In all three cases, we observe that the Non-Newtonian behavior of the gel dominates the impact dynamics where stresses and shear rates are low, either locally or globally. This is most clearly demonstrated by the transition from the Newtonian-like behavior of splashing (at low concentration, large size, or high velocity) to the yield stress behavior of sticking (at high concentration, small size, or low velocity). Each splash-to-stick transition shows impact characteristics of yield-stress fluids that are not observed in Newtonian droplets, including unique angles and curvatures in the ejection sheets, halted motion of the sheet contact line, thread-like ejected droplets, and contoured final surfaces.

\end{document}